\documentstyle[12pt]{article}
\begin{document}

\title{Compressible dynamics of magnetic field lines 
for incompressible MHD flows} 
\author{E.A. Kuznetsov$^{(a)}$\footnote{e-mail of the corresponding author:
kuznetso@itp.ac.ru}, 
T. Passot $^{(b)}$ and P.L. Sulem $^{(b)}$ \\ 
{\small $^{(a)}$ - \it Landau Institute for Theoretical Physics,  
2 Kosygin str., 119334 Moscow,  Russia }\\ 
{\small $^{(b)}$ - \it CNRS, Observatoire de la Cote d'Azur, 
PB 4229, 06304 Nice Cedex 4,  
France}} 
 
\date{} 
 
\maketitle 
 
\begin{abstract} 
It is demonstrated that the deformation of magnetic field  
lines in incompressible magnetohydrodynamic flows results from a   
compressible mapping. 
Appearance of zeroes for the mapping Jacobian correspond to the breaking of magnetic  field  
lines, associated with local blowup of the magnetic field. The possibility  
of such events is found to be unlikely in two dimensions  
but not in three dimensions.   
 
\end{abstract} 
 
\medskip 
 
 
\medskip 
 
\section{Introduction} 
An important property of ideal magnetohydrodynamics (MHD) is the  
frozenness of magnetic field in the plasma: 
fluid  particles remain pasted on  their  magnetic  
lines that are driven by the transverse  
velocity component. This remark is the  
starting point of a 
mixed Lagrangian-Eulerian description of ideal 
MHD flows, named  magnetic line  
representation (MLR) and first formulated   in \cite{mhd}. 
The idea originates from the vortex line representation (VLR) 
of hydrodynamic flows \cite{KR98} that involves  
a two-dimensional Lagrangian marker labeling each vortex  
line, together with a parameterization of these  lines. 
In three dimensions (3D), this representation enables  one to partially integrate 
the Euler equations  with respect to  
a continuous infinity of integrals of motion called the Cauchy invariants.   
 A main  
peculiarity of the transformation associated with the 
vortex line dynamics is its compressible character 
that, as recently pointed 
out by one of the authors \cite{0},    
is amenable of a simple  interpretation. The Euler equations can be rewritten as  
the equations of motion for a charged {\it compressible} fluid moving under  
the action of effective self-consistent electric and magnetic  fields 
satisfying  Maxwell equations.  
The new velocity coincides with the  velocity component transverse
to vorticity, which, due to the frozenness property, identifies with
the vortex line velocity.   
It is well known that the appearance of singularities in compressible flows  
is connected with the emergence  of shocks, corresponding to  
the  formation of folds in the classical catastrophe theory  
\cite{arnold}.  
In the gas-dynamic case, this process is completely characterized by the mapping defined by 
the transition from the usual 
Eulerian to the Lagrangian description.    
A zero of the Jacobian  
corresponds to the emergence  
of a singularity for the spatial derivatives of the velocity and density of  
the fluid.   
Due to the compressible character of VLR, the phenomenon  
of breaking becomes also possible for vortex lines in ideal incompressible fluids. 
Vortex line  breaking was first studied for three-dimensional 
integrable hydrodynamics  with Hamiltonian  
${\cal H}=\int |{\bf \Omega}|d{\bf r}$ where ${\bf \Omega}$ is the
vorticity \cite{10}.  
This model and the Euler equation are  
both incompressible and have the same symplectic operator  
defining the Poisson structure. Breaking  
of vortex lines is associated with the touching of two  
vortex lines and  results in an 
infinite vorticity.  
Recent  numerical simulations \cite{11,12} have suggested the possibility of 
such a scenario for the 3D Euler equations,  
but further investigations are required to reach a definite  
conclusion. 
In ideal MHD, we can expect the same behavior for the magnetic field 
which is a frozen-in quantity. 
In two dimensions (2D) however, the fact that vorticity is perpendicular to the  
flow plane while the magnetic field lies in it, puts a limit to the 
analogy, making magnetic field line breaking 
a  priori possible in  two dimensions, while singularities are excluded 
in 2D Euler flows. It will nevertheless be argued in this paper that 
magnetic field blowup is unlikely in 2D MHD.  
 
In Section 2, we recall the Cauchy formula for MHD flows, which 
plays a central role in the derivation of the Weber type transformation 
discussed in Section 3. This transformation is obtained  by  
extending  ideas of paper \cite{0} to ideal  
incompressible MHD flows. We in particular  indicate how  
the MHD equations can be partially integrated. Section 4 addresses  
the two-dimensional case where two conservation laws are established. 
In Section 5, we discuss the possibility of magnetic line 
breaking  as  a local blowup of the magnetic field,  
a process different from the gradient singularity associated with  
current sheets formation (\cite{parker} and references therein).  
A brief conclusion is provided by Section 6.
  
\section{Cauchy formula in MHD} 
 
 As well known, the magnetic field ${\bf h}$ in  ideal incompressible MHD 
obeys  
\begin{equation} 
\label{MHD} 
{\bf h}_t=\mbox{curl}({\bf v}\times {\bf h}), \,\,\, \mbox{div}\,{\bf v}=0,   
\end{equation} 
that formally coincides  with the equation governing 
the vorticity ${\bf \Omega}$ in  Euler hydrodynamics. 
Since only the transverse velocity ${\bf v}_{\perp}$ to the local
magnetic field is relevant in this equation, we  introduce  
new Lagrangian trajectories  
\begin{equation} 
\label{mapping} 
{\bf r}={\bf r}({\bf a},t), 
\end{equation} 
defined by  
\begin{eqnarray} 
\label{motion} 
&&\frac{d {\bf r}}{d t}={\bf v}_{\perp}({\bf r},t)\\ 
&&{\bf r}|_{t=0}={\bf a}.
\end{eqnarray}  
It is easily established that the Jacobian matrix 
(of element $\displaystyle{\hat J_{ij}=\frac{\partial x_j}{\partial a_i}}$) 
obeys 
\begin{equation} 
\label{jacoby} 
\frac{d}{dt}\hat J=\hat JU 
\end{equation} 
where the matrix $U$ has elements  
$\displaystyle{U_{ij}=\frac{\partial v_{\perp j}}{\partial x_i}}$. 
 One then obtains the equations for the Jacobian  
$J=\mbox{det}\,\hat J$ 
and for the inverse matrix $\hat J^{-1}$ 
with elements ${\partial a_j}/{\partial x_i},$  
(where ${\bf a}={\bf a(r},t)$ is the inverse of the   
mapping defined in (\ref{mapping})), in the form  
\begin{equation} 
\label{jacobian} 
\frac{d}{dt}J=  \mbox{div}\,{\bf v}_{\perp} J 
\end{equation} 
and  
\begin{equation} 
\label{inverse} 
\frac{d}{dt}\hat J^{-1}=-U\hat J^{-1}. 
\end{equation} 
Since $\mbox{div}\,{\bf v}_{\perp}$ is generically non zero, the mapping 
(\ref{mapping}) is compressible and the Jacobian $J$  can  
vanish. This observation is central in the discussion of the possibility 
of magnetic field blowup presented in Section 5. 
 
By means of eqs.  (\ref{jacobian}) and (\ref{inverse}), eq.   
(\ref{MHD}) is transformed into 
\begin{equation} 
\label{integral}  
D_t\left( Jh_{i}\frac{\partial a_j}{\partial x_i}\right)=0, 
\end{equation} 
where  
$D_t=\partial_t+({\bf v_{\perp}}\cdot\nabla)$ 
identifies with the material derivative  $d/dt$ used in (\ref{motion}). 
Integration of this equation leads to a ``new'' vector Lagrangian invariant 
\begin{equation} 
\label{invariant} 
 I_j({\bf a}) \equiv J\,h_{i}\frac{\partial a_j}{\partial x_i}  
\end{equation} 
that coincides with the initial magnetic field  
${\bf h}_0({\bf a})$
and is the analog of the Cauchy invariants 
for ideal hydrodynamics. The  magnetic field ${\bf h}$ is then given by  
\begin{equation} 
\label{field}  
{\bf h}({\bf r},t)=\frac{({\bf h}_0({\bf a})\cdot\nabla_{\bf a}){\bf r}({\bf a},t)}{J}. 
\end{equation}

\section{Weber type transformation} 
 
Equation  (\ref{field}) is the basis of the magnetic line representation  
\cite{mhd}. Another important formula for MLR follows from the  
velocity equation 
\begin{equation} 
\label{velocity} 
\partial_t{\bf v}+({\bf v}\cdot\nabla ){\bf v}=-\nabla p+ 
\mbox{curl}\,{\bf h}\times {\bf h},
\end{equation} 
where we normalized the magnetic field by the factor  
$\sqrt{4\pi\rho}$ 
(where $\rho$ is the density) so that ${\bf h}$ has the  dimension of  
a velocity. 
  
We also decompose the velocity ${\bf v}= {\bf v_{\perp}}+ {\bf v_{\tau}}$   
into transverse  and tangential  components 
and substitute in (\ref{velocity}). As a result,  
eq. (\ref{velocity}) is rewritten as 
\begin{equation} 
\label{electron} 
\partial_t {\bf v}_{\perp}+ 
({\bf v}_{\perp}\nabla){\bf v}_{\perp}= 
{\bf E}+{\bf v}_{\perp}\times {\bf H} + {\bf F^{\it mhd}}, 
\end{equation} 
where we introduced new  effective ``electric'' and ``magnetic'' fields  
\begin{equation} 
\label{electric} 
{\bf E}=-\nabla \left ( p+\frac{v^2_{\tau}}{2} \right )- 
\frac{\partial {\bf v}_{\tau}}{\partial t}, 
\end{equation} 
\begin{equation} 
\label{magnetic} 
{\bf H}=\mbox {\rm rot}~{\bf v}_{\tau}. 
\end{equation}  
In eq. (\ref{electron}), the force ${\bf F^{\it mhd}}={\bf j}\times {\bf h}$, 
involves the renormalized current 
\begin{equation} 
\label{current} 
{\bf j}=  
\mbox{curl}\,{\bf h}-{({\bf v}\cdot{\bf h})}/{h^2}\mbox{curl}\,{\bf v}. 
\end{equation} 
 
As seen from (\ref{electric}) and (\ref{magnetic}),  the 
new auxiliary  ``electric'' and ``magnetic''  fields  
can be expressed in terms of scalar and vector potentials  
$\displaystyle{\varphi= p+\frac{{\bf v}^2_{\tau}}{2}} $ and   
$\displaystyle{{\bf A}={\bf v}_{\tau}}$,  
so that the two Maxwell equations  
\begin{displaymath} 
\mbox {\rm div}~{\bf H}=0, \qquad 
\frac{\partial {\bf H}}{\partial t}= 
-\mbox {\rm curl}~{\bf E} 
\end{displaymath} 
are automatically satisfied. 
In this case, the vector potential ${\bf A}$  
has the gauge  
$$ 
\mbox{div}~ {\bf A}=- \mbox{div}~{\bf v}_{\perp}, 
$$ 
which is equivalent to the incompressibility 
condition $\mbox{div}~{\bf v}=0$. 
 
The two other Maxwell equations define  
auxiliary charge density  and current   
which follow from relations   
(\ref{electric}) and (\ref{magnetic}). 
 
New terms in the right hand side of eq. (\ref{electron}) 
also have  a mechanical interpretation. The Lorentz force  
${\bf v}_\perp\times {\bf H}$ plays the role of a  
Coriolis force. The potential $\varphi$  has a 
direct connection with the Bernoulli formula.  
The term ${\partial_t{\bf v_{\tau}}}$ results from the    
non-inertial character of the coordinate system.
 
In eq.  (\ref{electron}), we make the change of variable  
defined by mapping (\ref{mapping}). As a result,  
the equations of motion are expressed in a quasi-Hamiltonian form,  
analogous to eq. (20) of  \cite{0} \footnote{The first equation of the system  
(\ref{ham}) contains 
an addition term ${\bf F^{\it mhd}}$ and therefore we qualify  (\ref{ham})  
of quasi-Hamiltonian.} 
\begin{equation} 
\label{ham} 
D_t{\bf P}=-\frac{\partial h}{\partial {\bf r}}+{\bf F^{\it mhd}}, \qquad 
D_t{\bf r}=\frac{\partial h}{\partial {\bf P}}, 
\end{equation} 
where the  
Hamiltonian $h$  is given  
by the standard expression 
\begin{displaymath} 
h= \frac 12 ({\bf P}- {\bf A})^2 + \varphi \equiv p+\frac{{\bf v}^2}{2}, 
\end{displaymath} 
in terms of  
the generalized momentum 
${\bf P}={\bf v}_\perp +{\bf A}$ (that identifies  
with ${\bf v}$), and thus  
coincides with the Bernoulli "invariant" for a non-magnetic fluid. 
 
Introducing a new vector  
$$ 
u_k=P_i~\frac{\partial x_i}{\partial a_k}, 
$$ 
depending on $t$ and  ${\bf a}$,  one easily obtains from (\ref{ham})  
that this vector obeys 
\begin{equation} 
\label{u} 
D_t u_k=\frac{\partial}{\partial a_k} 
\left ( -p+\frac {v^2_{\perp}}{2}- \frac {v^2_{\tau}}{2}\right ) + 
 F^{\it mhd}_{i}~\frac{\partial x_i}{\partial a_k}. 
\end{equation} 
Using (\ref{field}) and the identity  
\begin{equation} 
\label{identity} 
\epsilon_{\alpha\beta\gamma}\frac{\partial x_i}{\partial a_{\beta}} 
\frac{\partial x_j}{\partial a_{\gamma}}=\epsilon_{ijk}J\, 
\frac{\partial a_{\alpha}}{\partial x_k}, 
\end{equation} 
one has  
$$ 
 F^{\it mhd}_{i}\frac{\partial x_i}{\partial {\bf a}}= 
{\bf h_0(a)}\times{\bf S}, 
$$ 
where  
$$ 
{\bf S}= ({\bf j}\cdot\nabla_r){\bf a}. 
$$ 
Equation  (\ref{u}) thus rewrites  
\begin{equation} 
\label{motion-u} 
D_t{\bf u}= \nabla_a\left ( -p+\frac {v^2_{\perp}}{2}-  
\frac {v^2_{\tau}}{2}\right ) + 
{\bf h_0(a)}\times{\bf S}. 
\end{equation} 
 
Integrating in time then leads to the Weber type  
transformation 
\begin{equation} 
\label{weber} 
{\bf u}= {\bf u_0(a)} +\nabla_a \Phi+ 
{\bf h_0(a)}\times{\bf W}, 
\end{equation} 
where the potential $\Phi$ satisfies a Bernoulli type equation, 
$$ 
D_t\Phi= -p+\frac {v^2_{\perp}}{2}- \frac {v^2_{\tau}}{2} 
$$ 
and the vector ${\bf W}$ obeys  
\begin{equation} 
\label{W} 
D_t{\bf W}={\bf S}. 
\end{equation} 
 
If initially $\Phi|_{t=0}=0$ and  ${\bf W}|_{t=0}=0$, the integration  
``constant'' ${\bf u}_0({\bf a})$  coincides with the initial velocity 
${\bf v}_0({\bf a})$.  This  vector ${\bf u}_0({\bf a})$ 
is thus a new Lagrangian invariant.

To get a closed description we eliminate the pressure $p$ by applying 
the  curl operator (with respect to ${\bf a}$-variables) 
on eq.  (\ref{weber}) 
\begin{equation} 
\label{1weber} 
\mbox{curl}_a \,{\bf u}=\mbox{curl}_a \, {\bf u_0(a)} 
+\mbox{curl}_a \, [{\bf h_0(a)}\times{\bf W}]. 
\end{equation} 
This equation can  also be rewritten as 
\begin{equation} 
\label{2weber} 
{\bf\Omega(r,}t)=\frac{({\bf\Omega_0(a},t)\cdot \nabla_a){\bf r(a},t)}{J}. 
\end{equation} 
Here ${\bf\Omega_0(a,}t)$ is given by 
$$ 
{\bf\Omega_0(a,}t)= {\bf\Omega_0(a})+\mbox{curl}_a \, [{\bf h_0(a)} 
\times{\bf W}], 
$$ 
where ${\bf\Omega_0(a})$ is the initial vorticity.  
When  ${\bf h_0(a) }=0$, eq. (\ref{2weber}) 
reduces to the Cauchy formula for vorticity in ideal hydrodynamics.

The vector ${\bf W}$  is  
determined from  eq. (\ref{W}) that rewrites 
\begin{equation} 
\label{W1} 
D_t{\bf W}=\frac{({\bf v}\cdot{\bf b})}{b^2}~{\bf \Omega}_0({\bf a},t)- 
\frac 1J\,  
\mbox{curl}_a\,\left(\frac{\hat g\,{\bf h}_0({\bf a})}{J}\right) 
\end{equation} 
where $\hat g$ is the MLR metric tensor defined by  
$$ 
g_{\alpha\beta}=\frac{\partial x_i}{\partial a_{\alpha}}\cdot 
\frac{\partial x_i}{\partial a_{\beta}} 
$$ 
and ${\bf b}= J{\bf h}$ is given by  (\ref{field}). 
  
As a result, we have two equations of motion for the mapping (\ref{motion})  
and for the vector ${\bf W}$. Together with eqs.
(\ref{field}),  (\ref{2weber}) and the relation between velocity and 
vorticity, 
\begin{equation} 
\label{connection}  
{\bf \Omega}=\mbox{curl}_r{\bf v},\,\,\,\, \mbox{div}_r\,{\bf v}=0,   
\end{equation} 
this constitutes a closed system of equations that provides a magnetic line 
representation for incompressible MHD  (to be compared with 
\cite{mhd}). These equations are solved with respect to two Lagrangian  
invariants 
${\bf h}_0({\bf a})$ and  ${\bf \Omega}_0({\bf a})$.  
It is possible to show \cite{mhd} that conservation of these invariants in  
MHD is  
a consequence of  relabeling symmetry,  
as it is the case for Euler equation (see, e.g. the reviews \cite{17,16}).   
 
The magnetic line representation involving the local 
change of variables ${\bf r}={\bf r}({\bf a},t)$, breaks down at singular  
points where the Jacobian is zero or infinity and the normal 
velocity is not defined.  
 
Let us consider the null  
point ${\bf r}={\bf r}(t)$  defined by  
\begin{equation} 
\label{zero} 
{\bf h}({\bf r}(t),t)=0.  
\end{equation} 
Differentiating this equation with respect to time, we get 
$$ 
\frac{\partial \bf h}{\partial t}+({\bf \dot r}(t)\cdot \nabla){\bf h}=0, 
$$ 
with ${\bf \dot r}(t) = {\bf v}({\bf r}(t), t)$, which
shows   that the null points are advected  
by the flow.  
The velocity ${\bf v}$ at these points is defined by inverting the  
curl operator in (\ref{connection}). 
 
Null-points are topological singularities for the tangent vector  
field $\tau({\bf r})$. 
Their classification depends  on  the space dimension 
$D$. Topological constraints  that 
can be considered as additional conditions 
for the MLR system, 
can be written as integrals of the vector field  
${\bf \tau}({\bf r})$ 
and its derivatives  
over the boundary of simply-connected regions (in 3D) or along closed  
contours (in 2D)  enclosing the null-points.  
In $D=2$, one has 
\begin{equation} 
\label{con2} 
\oint (\nabla \varphi\cdot d{\bf r}) =2\pi m,  
\end{equation} 
where $\varphi$ is the angle between the vector 
${\bf \tau}$ and the $x$-axis and  
$m$  is an integer often  called 
topological charge. It   
is equal to the total number of turns of the vector ${\bf \tau}$  
while passing around the closed contour encircling the null-point.  
For instance, for $X$-points or $O$-points, $m=\pm 1$.  
 
In $D=3$,  the topological charge is defined 
as the degree of the mapping ${\cal S}^2\to{\cal S}^2$, given by 
\begin{equation} 
\label{con3} 
\int_{\partial V}\epsilon_{\alpha\beta\gamma}~({\bf \tau}\cdot  
[\partial_{\beta} 
{\bf \tau}\times\partial_{\gamma}{\bf \tau} ])~dS_{\gamma}= 4\pi m, 
\end{equation} 
where the integration is performed  over the boundary  
$\partial V$ of a region $V$ containing null-points. 
 
Conditions (\ref{connection})- (\ref{con3}) 
complete the MLR equations in the general case when the Jacobian has  
localized zeroes. 
 
The above representation involves simultaneous use of Lagrangian  variables  in 
eqs. (\ref{motion}), (\ref{W1}), (\ref{field}), (\ref{2weber})  
and Eulerian ones in (\ref{connection}), making the numerical
integration of these equations very cumbersome. 
It is therefore of interest  
to look for a representation formulated in the sole physical space. 
 
Let us consider the inverse of the mapping  
${\bf a}={\bf a(r},t)$. 
Using eq. (\ref{motion}), one has  
\begin{equation} 
\label{invmotion}  
\partial_t{\bf a}+({\bf v_{\perp}}\cdot\nabla) {\bf a}=0. 
\end{equation}  
 From (\ref{identity}), eq.  (\ref{field}) for the magnetic field 
rewrites 
\begin{equation} 
\label{field1}  
{\bf h}=\epsilon_{ijk}h_{0i}({\bf a})[\nabla a_j\times\nabla a_k]. 
\end{equation} 
 
Formula  (\ref{2weber}) for the vorticity in ${\bf r}$-variable becomes 
\begin{equation} 
\label{3weber} 
{\bf\Omega(r,}t)=\mbox{curl}( V_i\nabla a_i) 
\end{equation} 
where 
$$ 
{\bf V}={\bf v_{0}}({\bf a})+{\bf h_0({\bf a})\times W}. 
$$ 
Similarly, the equation of motion (\ref{W}) for the vector ${\bf W}$   
transforms into 
\begin{equation} 
\label{W2}  
\partial_t{\bf W}+({\bf v_{\perp}}\cdot\nabla){\bf W}= 
 -({\bf j}\cdot\nabla){\bf a}, 
\end{equation} 
with initial condition ${\bf W}|_{t=0}=0$. 
Here the generalized current ${\bf j}$ is given by (\ref{current}).

These equations are completed by relation (\ref{connection}) and the 
definition of the normal velocity 
${\bf v}_{\perp}=\widehat\Pi~ {\bf v}$, 
where the projector $\widehat\Pi$ is defined by means of the unit tangent vector 
${\tau}={\bf h}/h$ as  
$\Pi_{\alpha\beta}=\delta_{\alpha\beta}-\tau_{\alpha}\tau_{\beta}$.  
They provide  a closed system for 
ideal MHD flows, where all the spatial derivatives are taken with  
respect to ${\bf r}$-variables.

\section{Conservation laws in two dimensions} 
 
The magnetic line representation significantly simplifies in two
dimensions where the magnetic field lies on the same plane as the flow.   
It is convenient to introduce, instead of the initial position ${\bf a}$, 
 the scalar magnetic potential $\psi$  
defined by  
$$ 
h_x=\frac{\partial\psi}{\partial y},\,\,\ h_y=-\frac{\partial\psi}{\partial x}, 
$$ 
and  a Cartesian coordinate $y$. 
 
 By fixing $\psi$, we select a magnetic line given by
$$ 
\frac{dx}{\partial\psi/\partial y}=-\frac{dy}{\partial\psi/\partial x}.
$$ 
The difference 
$\psi_1-\psi_2$ is equal to the flux of magnetic field between two lines 
with different values of $\psi$.  
 
In 2D, $\psi$ is a Lagrangian invariant,  since  it follows from the integration 
of the induction equation (\ref{MHD}) that 
\begin{equation} 
\label{psi1} 
\frac{\partial \psi}{\partial t}+({\bf v}\cdot\nabla)~\psi=0. 
\end{equation} 
The  potential  
\begin{equation} 
\label{psi}  
\psi=\psi(x,y,t) 
\end{equation}  
can then be taken as a Lagrangian marker of the magnetic lines. 
Solving locally eq.  (\ref{psi}) in the form  
$y=y(x,\psi,t)$, 
provides the desired mapping that replaces  (\ref{mapping}).    
 
This change of variables, being a mixed Lagrangian-Eulerian one,  
realizes a transformation  to a {\it curvilinear} system of coordinates 
movable with magnetic lines.  
In order to implement  the transformation from variables $(x,y,t)$ to  
$(x,\psi,t)$  in eqs. (\ref{psi1}) and (\ref{velocity}), 
we use  
\begin{eqnarray} 
&&\frac{\partial f}{\partial t}=\frac{1}{y_{\psi}}\, 
[f_ty_{\psi}-f_{\psi}y_t],\label{der1}\\ 
&&\frac{\partial f}{\partial x}= \frac{1}{y_{\psi}}\, 
[f_xy_{\psi}-f_{\psi}y_x],\label{der2}\\ 
&&\frac{\partial f}{\partial y}=\frac{f_{\psi}}{y_{\psi}}. \label{der3} 
\end{eqnarray}   
where derivatives are taken 
relatively to $(x,y,t)$ in the left hand sides of the above equations and to $(x,\psi,t)$  
in the right hand sides. 
 
Equation  (\ref{psi})  for the magnetic potential then  
transforms into an  equation for the  
magnetic line $\psi$ 
\begin{equation} 
\label{y} 
y_t+v_xy_x=v_y. 
\end{equation} 
This equation is a  kinematic condition. 
As the equation of motion (\ref{motion}), the 
dynamics of $y$ is prescribed by the  velocity component  
normal to the magnetic field line $y_t=v_{\perp}\sqrt{1+y_x^2}$ 
where $v_{\perp}=({\bf v}\cdot{\bf n})$ and  
$\displaystyle{{\bf n}=\frac{1}{\sqrt{1+y_x^2}} (-y_x, 1)}$. 
In terms of the new variables, the magnetic field is given by 
$$ 
h_x=\frac{1}{y_{\psi}},\,\,\, h_y=\frac{y_x}{y_{\psi}}, 
$$ 
which are equivalent to the Cauchy formula (\ref{field}) for the magnetic field 
in  2D. The derivative $y_{\psi}$ in the denominators holds for  
the Jacobian $J$.  
The  equation for the quantity $y_{\psi}$ can be found by  differentiating  
(\ref{y}) with respect to $\psi$ and applying  
 the  incompressibility condition in the form 
\begin{equation} 
\label{incom}  
\frac{\partial v_x}{\partial x}~ y_{\psi}-  
\frac{\partial v_x}{\partial \psi}~ y_{x} 
+\frac{\partial v_y}{\partial \psi}=0. 
\end{equation} 
This results in a continuity equation for $y_{\psi}$, 
\begin{equation} 
\label{contin} 
\partial_ty_{\psi}+ \partial_ x(v_x y_{\psi})=0, 
\end{equation} 
so that $y_{\psi}$ has the meaning of a layer density.   
 
Another useful relation can be obtained from the equations 
for the velocity  components $v_x$ and $v_y$ that now read 
\begin{equation} 
\label{v-x} 
\partial_tv_x+ v_x\partial_ x v_x =  
-\partial_xp+(\partial_{\psi}p-j)\frac{y_x}{y_{\psi}}, 
\end{equation} 
\begin{equation} 
\label{v-y} 
\partial_tv_y+ v_x\partial_ x v_y=-(\partial_{\psi}p-j)\frac{1}{y_{\psi}}, 
\end{equation} 
where $j=\mbox{curl}~{\bf h}$ is the current directed along the $z$ direction. 
 It is then convenient to introduce 
$$ 
U=v_x+y_x v_y, 
$$ 
where  $y_x$ obeys the equation  
$$ 
\partial_ty_x+v_x\partial_ xy_x +y_x \partial_xv_x = \partial_xv_y 
$$ 
derived from (\ref{y}). 
The function $U$  coincides up to the factor $1/\sqrt{1+y_x^2}$ with the  
velocity component tangent to the magnetic field 
$\displaystyle{v_{\tau}= \frac{1}{\sqrt{1+y_x^2}}U}$. One easily gets  
\begin{equation} 
\label{U} 
\partial_tU+  \partial_x(v_x U)=-\partial_x(p-v^2/2), 
\end{equation}  
that can be viewed as a differential form of the Kelvin theorem. 
 
Combination of eqs. (\ref{contin}) and (\ref{v-y}) gives that
$w=v_yy_{\psi}$ obeys
\begin{equation} 
\label{w} 
\partial_tw+\partial_x(v_x w)=-\partial_{\psi}p +j. 
\end{equation} 
To find the analog of (\ref{2weber}) in the 2D case, it is  
convenient to make the 
change of variables $y=y(x,\psi,t)$ in the vorticity equation 
\begin{equation} 
\label{2vorticity}   
\partial_t \Omega +({\bf v}\cdot \nabla)\Omega=\nabla j\times 
\nabla \psi. 
\end{equation} 
Substituting  relations  (\ref{der1}- \ref{der3}) into (\ref{2vorticity}) and 
using eq.  (\ref{y}), we get 
$$ 
\partial_t\Omega +v_x\partial_x\Omega_x=\frac{\partial_xj}{y_{\psi}}. 
$$

Equations (\ref{contin}) and (\ref{U}) provide conservation laws for 
2D incompressible MHD. They remain valid in the hydrodynamic limit,  
provided $\psi$ is replaced by vorticity  or by any other Lagrangian invariant.

\section{Possibility of magnetic line breaking } 
 
An important property of the magnetic line representation  
concerns  the compressibility 
of the mapping (\ref{mapping}), which permits  
magnetic line breaking. 
At  the breaking point, the magnetic field, according to  (\ref{field}),  
becomes infinite due to the vanishing of the Jacobian.  
As it follows from references  \cite{0,10,11,12}, the   
possibility of vortex line  breaking  depends on the space dimension.  
For two-dimensional flows described by the Euler equations, 
vorticity is perpendicular to the flow plane 
and therefore $\mbox{div}\,{\bf v_{\perp}}=0$. As the consequence, the  
corresponding mapping is incompressible  and the Jacobian remains constant.  
  
For 2D incompressible MHD, the situation  is different since the   
magnetic field  
lies in  the flow plane. 
The velocity can therefore be decomposed into 
transverse and longitudinal components relative to {\it the magnetic  
field direction}.  
In such a case $\mbox{div}\,{\bf v_{\perp}}\neq 0$ and 
the breaking of magnetic lines is not a priori excluded.  
 
Let us thus assume that a breaking of magnetic lines occurs. Denote
by  $t=\tilde t({\bf a})>0$ the positive roots of the equation  
$$ 
J({\bf a}, t)=0 ,
$$  
 and find the minimal value $t_0=\min_a\,\tilde t({\bf a})$ 
which defines the first instant of time when the Jacobian vanishes. Let  
${\bf a=a_0}$ 
be the Lagrangian coordinate of the point where this minimum is attained.   
We first consider that near  
the singular point, as $t\to t_0$,  
the Jacobian behaves as  
\begin{equation} 
\label{expansion} 
J=\alpha(t_0-t)+ \gamma_{ij}\Delta a_i\Delta a_j  
\end{equation}  
where $\alpha>0$, $\gamma_{ij}$ is a positive definite  
(generically non-degenerated) matrix and  
$\Delta{\bf a}={\bf a -a}_0$. This assumes that the magnetic field
does not vanish at the collapse point and in particular that  
the three vectors $\partial{\bf r}/\partial a_i$  
($i=1,2,3$) 
lie in the same plane, with none of them vanishing.  
In this case, eq. (\ref{field}) rewrites  
\begin{equation} 
\label{sing} 
{\bf h}=\frac{\bf b}{\alpha(t_0-t)+ \gamma_{ij}\Delta a_i\Delta a_j}. 
\end{equation}  
where ${\bf b}= ({\bf h_0}(\bf a)\cdot\nabla_a){\bf r}|_{t_0,a_0}$. 
This corresponds to a blowup of the magnetic 
field ${\bf h}({\bf a}_0)$  like $1/(t_0-t)$.

The MHD equations conserve the energy ${\cal E}$ given by the sum of  
the kinetic 
$\displaystyle{{\cal E}_k =\int\frac{\bf v^2}{2} d{\bf r}} $  
and magnetic   
$\displaystyle{{\cal E}_h = \int\frac{\bf h^2}{2} d{\bf r} }$ energies, 
that  both have to remain finite as $t\to t_0$ .  
 
Let us the estimate the 
contribution provided by a possible singularity (\ref{sing}), to the magnetic energy 
\begin{equation} 
\label{contr} 
{\cal E}_h \approx\int\frac{b^2}{J^2}d{\bf r}. 
\end{equation} 
By changing variables from ${\bf r}$ to ${\bf a}$, the contribution to  
this integral arising from a ball of radius  $R\sim \tau^{1/2}$ 
where $\tau=t_0-t$ and centered in ${\bf a}_0$, rewrites  
\begin{equation} 
\label{int} 
{\cal E}_h^s \approx b^2\int\frac{d{\bf a}}{\alpha\tau+ \gamma_{ij}a_ia_j} 
\propto  (t_0-t)^{(D-2)/2}. 
\end{equation} 
 
The size of the retained ball is the largest compatible with the asymptotics. 
The contribution from the other region being most likely finite, we conclude 
that a magnetic field blowup in not excluded in 3D for the assumed expansion 
of the Jacobian. The same conclusion holds if the Jacobian vanishes like 
$(t_0-t)^n$ at the singularity point, with a ball size modified accordingly. 
At a point where the matrix $\gamma$ is degenerated with e.g. one   
eigenvalue $\lambda_1$ being zero, the Jacobian locally  
becomes  
\begin{equation} 
\label{deg} 
J=\alpha(t_0-t)+ \tilde\gamma_{ij}a^{\perp}_ia_j^{\perp} +\beta a_1^4. 
\end{equation} 
where ${\bf a}^{\perp}$ holds for the projection of the vector ${\bf a}$, 
transverse to the direction 
of the eigenvector associated with  the zero eigenvalue. 
The contribution of the singularity to the magnetic energy then scales like  
${\cal E}_h^s \sim (t_0-t)^{1/4}$, a behavior which again does not contradict 
the possible existence of a singularity. 
 
In $D=2$, the conclusion can be different. Since the contribution of the 
selected ball 
to the magnetic energy does not tend to zero as $t \to t_0$,  
a small extension of this domain  to a ball of  size $R$ can lead to  
a logarithmic divergence 
${\cal E}_h^s \sim B^2 \log \frac{\gamma R^2}{\alpha \tau}\to \infty $. 
The divergence becomes more dramatic in the case of a degenerate matrix 
$\gamma$, for which  
${\cal E}_h^s \sim (t_0-t)^{-1/4}$. 
This observation leads us to conjecture that a blowup of the
magnetic field is probably excluded in two dimensions but not 
necessary in three dimensions. Note that the conservations laws (\ref{contin}) 
and (\ref{U}) for the two-dimensional problem derived in Section 3, could  
possibly be useful for a rigorous proof of the absence of magnetic blowup. 
 
\section{Conclusion} 
 
The mechanism for a finite-time singularity addressed in this paper  
corresponds to the breaking of magnetic field lines resulting in a  
catastrophic amplification of the local magnetic field strength.  
It is worth to notice that this process does not contradict  
the necessary condition for blowup in MHD \cite{CKS} that represents  
the analog of the Beale-Kato-Majda inequality \cite{BKM}. According to  
\cite{CKS} the velocity and magnetic field  
retain their smoothness on a time interval $[0,T]$ as long as  
$$ 
\int^T_0 (|\Omega(t)|_{\infty}+ |j(t)|_{\infty})dt<\infty.  
$$ 
Hence a finite-time singularity  of any kind must be accompanied by the  
blow-up  
of $\Omega$ and ${\bf \nabla h}$. However,  
this criterion does not exclude a blowup of the magnetic field as well.  
Constraints are nevertheless provided by regularity theorems;   
a result for example  states that  
the solution remains globally smooth  if   
the initial magnetic field has a mean component 
 sufficiently large  
compared to the fluctuations, assumed to be localized \cite{sulems}. 
This property is a consequence of the fact that only  
counter-propagating Alfv\'en wave packets nonlinearly interact.  
 
A specific conclusion of this paper is that magnetic  
field blowup resulting from magnetic line breaking is unlikely 
in two dimensions. Nevertheless, the  
present formalism cannot capture the behavior near a neutral X-point.  
Numerical evidence and self-similar reductions however indicate that  
in this case the 
current amplification is exponential in time \cite{FPSM} \cite{SFPM}. 
 
Furthermore, recent direct numerical simulations of 3D MHD indicate  
the formation of  
quasi-two dimensional current sheets that result in a depletion of the 
nonlinearity strength \cite{grauer}, a mechanism that could prevent 
singularities. In order to validate the blowup scenario discussed in
this paper, it is thus of interest to look for initial conditions 
that do not lead to bidimensionalization and has an  
initial velocity field whose component transverse to the local  
magnetic field has a significant divergence. 
   
\section*{ Acknowledgments} 
 
This work was supported by INTAS (grant no. 00-00292). 
The work of E.K. was 
also supported by the RFBR (grant no. 00-01-00929).  
E.K. wishes to thank the  Observatoire de la  
C\^ote d'Azur, where this work was initiated and completed,  
for its kind hospitality during visits supported by the Landau-CNRS agreement.

\end{document}